\documentclass[aps,prl,twocolumn,amsmath,amsart,amssymb,superscriptaddress]{revtex4}
\usepackage{graphicx}
\usepackage[T1]{fontenc}
\usepackage{amsmath}
\usepackage{color}
\usepackage{times}

\usepackage{ulem}

\newcommand{\BFCA}{Ba(Fe$_{1-x}$Co$_x)_2$As$_2$}
\newcommand{\BFA}{BaFe$_2$As$_2$}

\begin{document}

\title{Spin-excitation anisotropy in the nematic state of detwinned FeSe}

\author{Xingye Lu}
\email{luxy@bnu.edu.cn}
\affiliation{Center for Advanced Quantum Studies, Applied Optics Beijing Area Major Laboratory, and Department of Physics, Beijing Normal University, Beijing 100875, China}

\author{Wenliang Zhang}
\author{Yi Tseng}
\affiliation{Photon Science Division, Swiss Light Source, Paul Scherrer Institut, CH-5232 Villigen PSI, Switzerland}

\author{Ruixian Liu}
\author{Zhen Tao}
\affiliation{Center for Advanced Quantum Studies and Department of Physics, Beijing Normal University, Beijing 100875, China}

\author{Eugenio Paris}
\affiliation{Photon Science Division, Swiss Light Source, Paul Scherrer Institut, CH-5232 Villigen PSI, Switzerland}

\author{Panpan Liu}
\affiliation{Center for Advanced Quantum Studies and Department of Physics, Beijing Normal University, Beijing 100875, China}

\author{Tong Chen}
\affiliation{Department of Physics and Astronomy, Rice Center for Quantum Materials, Rice University, Houston, TX 77005, USA}

\author{Vladimir Strocov}
\affiliation{Photon Science Division, Swiss Light Source, Paul Scherrer Institut, CH-5232 Villigen PSI, Switzerland}

\author{Yu Song}
\affiliation{Department of Physics, University of California, Berkeley, California 94720, USA}

\author{Rong Yu}
\affiliation{Department of Physics, Renmin University of China, Beijing 100872, China}

\author{Qimiao Si}
\author{Pengcheng Dai}
\email{pdai@rice.edu}

\affiliation{Department of Physics and Astronomy, Rice Center for Quantum Materials, Rice University, Houston, TX 77005, USA}

\author{Thorsten Schmitt}
\email{thorsten.schmitt@psi.ch}
\affiliation{Photon Science Division, Swiss Light Source, Paul Scherrer Institut, CH-5232 Villigen PSI, Switzerland}

\date{\today}

\begin{abstract}

The origin of the electronic nematicity in FeSe is one of the most important unresolved puzzles in the study of iron-based superconductors. In both spin- and orbital-nematic models, the intrinsic magnetic excitations at $\mathbf{Q}_1=(1, 0)$ and $\mathbf{Q}_2=(0, 1)$ of twin-free FeSe are expected to provide decisive criteria for clarifying this issue. Although a spin-fluctuation anisotropy below 10 meV between $\mathbf{Q}_1$ and $\mathbf{Q}_2$ has been observed by inelastic neutron scattering around $T_c\sim 9$ K  ($<<T_s\sim 90$ K), it remains unclear whether such an anisotropy also persists at higher energies and associates with the nematic transition $T_{\rm s}$. Here we use resonant inelastic x-ray scattering (RIXS) to probe the high-energy magnetic excitations of uniaxial-strain detwinned FeSe and {\BFA}. A prominent anisotropy between the magnetic excitations along the $H$ and $K$ directions is found to persist to $\sim200$ meV in FeSe, which is
even more pronounced than the anisotropy of spin waves in {\BFA}. This anisotropy decreases gradually with increasing temperature and finally vanishes at a temperature around the nematic transition temperature $T_{\rm s}$. Our results reveal an unprecedented strong spin-excitation anisotropy with a large energy scale well above the $d_{xz}/d_{yz}$ orbital splitting, suggesting that the nematic phase transition is primarily spin-driven.
Moreover, the measured high-energy spin excitations are dispersive and underdamped, which can be understood from a local-moment perspective. 
Our findings provide the much-needed understanding 
of the mechanism for the nematicity
of FeSe and points to a unified description of the correlation physics across seemingly
distinct classes of Fe-based superconductors.

\end{abstract}

\maketitle

Intertwined order and fluctuations in high transition-temperature ($T_{\rm c}$) superconductors are pivotal for understanding the microscopic origin of superconducting electron pairing \cite{fradkin15}. Of particular interest is the electronic nematic state present in both cuprate and iron-based superconductors (FeSCs). Initially discovered through in-plane electronic anisotropy with $C_2$ symmetry in the paramagnetic orthorhombic state of detwinned {\BFCA} \cite{JHChu2010, yi2011, JHChu2012}, the electronic nematic state and its fluctuations has been identified as a ubiquitous feature of FeSCs and is believed to be essential for the structural and magnetic transitions in
FeSCs \cite{Fernandes14, NatRevMat16,Kuo2016, bohmer16, bohmer18},
and may enhance the electron pairing for high-$T_{\rm c}$ superconductivity \cite{senthil, kivelson, paglione20}.

\begin{figure*}
\includegraphics[width=15cm]{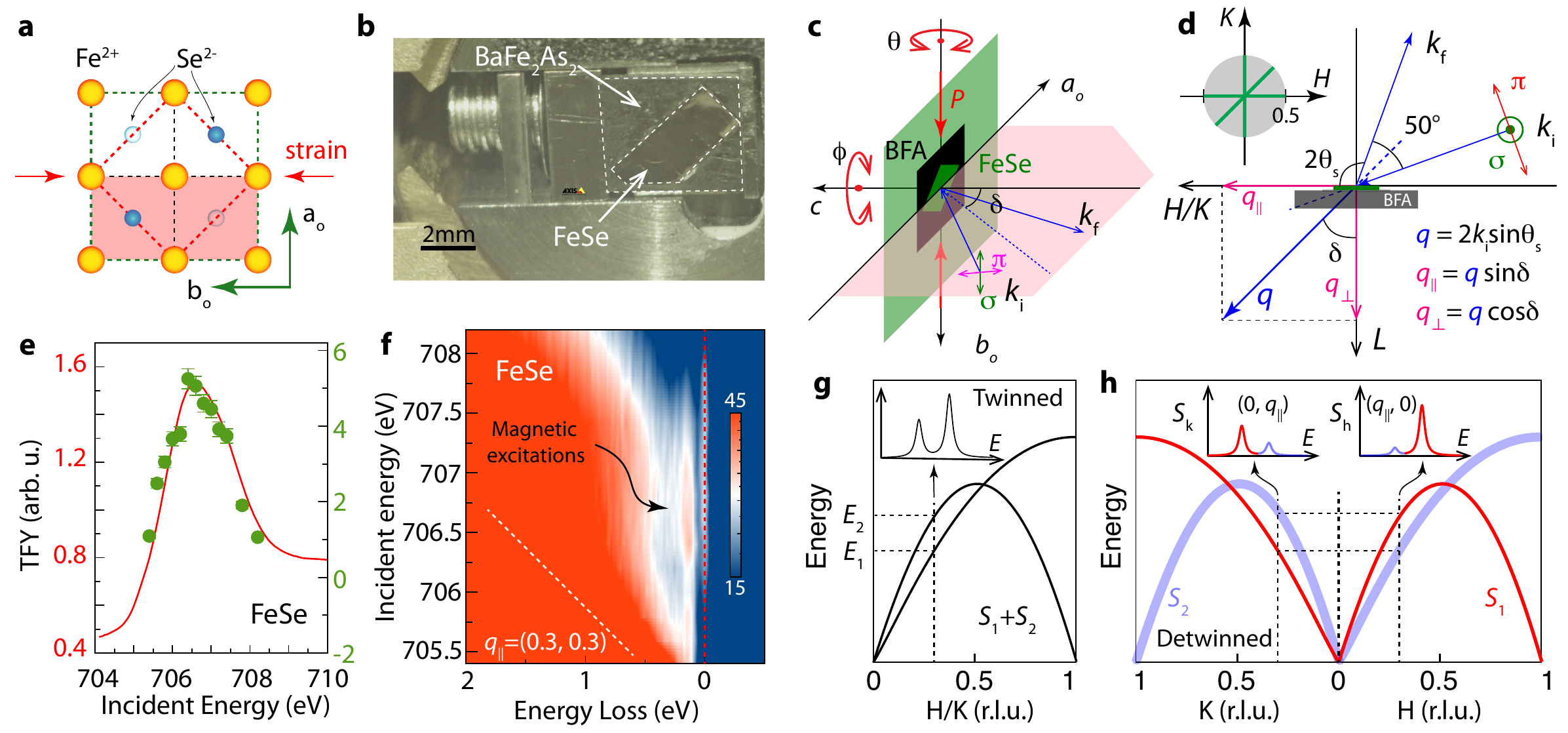}
\caption{Crystal structure, detwinning strategy, scattering geometry and incident energy dependent RIXS. {\bf a,} Structure of FeSe layer. FeSe crystal consists of stacked FeSe layers. The yellow filled circles mark Fe$^{2+}$ ions. The filled and open blue circles denote Se$^{2-}$ ions above and below Fe-Fe plane, respectively. The horizontal arrows mark the direction of uniaxial strain. The red dashed diamond and green dashed square denote the tetragonal and orthorhombic unit cell, respectively. {\bf b,} A mechanical detwinning device installed on the RIXS spectrometer, with a thin FeSe crystal glued on a pre-cleaved {\BFA} crystal that is pressured in the device. {\bf c, d,} Scattering geometry for RIXS measurements. The sample rotation $\theta$ around the vertical axis controls the in-plane momentum transfer $q_\shortparallel$ and the rotation $\phi$ around $c$-axis can tune the scattering plane (pink area). The gray filled circle marks the momentum area accessible in this study ($\mathbf{q}\lessapprox0.5$). The green lines in the gray area show the high-symmetry directions for RIXS measurements. {\bf e,} The red curve is the total fluorescence yield (TFY) XAS spectrum of FeSe collected near Fe-$L_3$ edge. The green dots give the integrated intensity of the magnetic excitations shown in {\bf f}. {\bf f,} Incident-energy dependence of the excitations of FeSe at $q_\shortparallel=(0.3, 0.3)$, measured near the Fe-$L_3$ edge with $\pi$ polarization at $T=20$ K. The magnetic excitations are marked by a curved arrow. {\bf g, h,} Schematics of magnetic excitation dispersions for twinned ({\bf g}) and detwinned ({\bf h}) FeSe assuming a difference between $S_2$ and $S_1$. Correspondingly, the insets in {\bf g} and {\bf h} show schematic RIXS energy spectra for twinned and detwinned FeSe.}
\label{fig1}
\end{figure*}

Iron selenide (FeSe) is a unique material among FeSCs because of its simplest structure (Fig. \ref{fig1}a) \cite{Hsu} and unusual electronic properties, such as strong anisotropy of the superconducting order parameter, extended electronic nematic phase, and highly tunable $T_{\rm c}$ \cite{bohmer18}. In particular, different from iron pnictides, which has a collinear antiferromagnetic (AF) ground state below the tetragonal-to-orthorhombic structural (nematic) phase transition $T_{\rm s}$, FeSe exhibits a similar nematic transition ($T_{\rm s}\approx 90$ K) \cite{McQueen09}, but has no static AF order, providing a broad temperature range below $T_{\rm s}$ as an ideal platform for investigating the electronic nematicity and its interplay with superconductivity.

FeSe consists of stacked charge-neutral FeSe layers (Fig. \ref{fig1}a). Upon cooling, it undergoes a nematic transition at $T_{\rm s}\approx 90$ K, below which twin domains form along two mutually perpendicular directions and exhibit macroscopic four-fold symmetry, impeding the study of the intrinsic electronic properties of the orthorhombic (nematic) state.
Through detwinning FeSe using uniaxial strain, resistivity and electronic structure measurements reveal strong electronic anisotropy in the nematic state \cite{tanatar16, Coldea18}. At lower temperature, FeSe enters a superconducting ground state with $T_c\approx9$ K, in which a superconducting energy gap anisotropy has been observed via angle resolved photoemission spectroscopy (ARPES) \cite{Hashimoto2018, Liu2018, Rhodes2018} and scanning tunnelling spectroscopy (STM) \cite{sprau17}.%

In the presence of orbital splitting (or orbital ordering) between $d_{xz}$ and $d_{yz}$ orbitals \cite{yi2019, Coldea18, CCLee2009}, and the absence of magnetic order, the electronic nematic phase was suggested to be driven by orbital fluctuations \cite{Baek15, anna15, Yamakawa16, Onari16}. On the other hand, various experimental evidences, in particular the discovery of intense magnetic excitations and their correlation with the nematic transition \cite{wang16, Qwang16, MWMa17}, emphasized the importance of the spin degree of freedom in driving the electronic nematic order. In addition, various localized models based on quantum paramagnetism, spin frustration and magnetic quadrupolar order were also proposed to account for the nematic transition, the
magnetic excitations, and the absence of AF order \cite{FWang15, YuSi15,Glasbrenner15, She17}.

It is proposed that resolving the intrinsic magnetic excitations in twin-free FeSe 
is the key for clarifying the microscopic origin of the unusual electronic anisotropy in both the superconducting and nematic 
state \cite{Fernandes14,Lu14,NatRevMat16}.
Through employing {\BFA} as substrate for applying uniaxial strain, some of us have recently measured the low-energy spin fluctuations ($E\lessapprox10$ meV) of detwinned FeSe using inelastic neutron scattering \cite{Chen19}. These results have revealed anisotropic spin fluctuations in the normal state and a spin resonance appearing at only ${\bf Q}_1=(1, 0)$ below $T_{\rm c}$, consistent with the picture of orbital-selective Cooper
pairing \cite{sprau17,Yu14, Fanfarillo2016}.
However, neutron scattering experiments were unable to determine what happens to the magnetic excitation anisotropy across the nematic transition due to enhanced background scattering from the large aluminum detwinning device on warming to $T_{\rm s}$.  In addition, the energy scale of the magnetic anisotropy is unknown because the background magnetic scattering from the {\BFA} substrate overwhelms the magnetic signal from FeSe for energies above 10 meV \cite{Chen19}.

An ideal method to probe the intrinsic magnetic excitations of FeSe is to use resonant inelastic x-ray scattering (RIXS) at the Fe-$L_3$ edge in combination with the above described detwinning method (Fig. \ref{fig1}b) \cite{kejin2013, johnny17, monney19, boothroyd19, johnny19, johnny21}. RIXS at transition-metal $L$ edges has been widely used to study the (para)magnons of cuprate and FeSCs, as well as various elementary excitations including phonon, crystal-field excitations and plasmons \cite{kejin2013, johnny17, monney19, boothroyd19, johnny19, johnny21, rixs_rmp, thorsten12, peng17, chunjing_prx, weisheng18}. Since Fe-$L_3$ X-ray ($707$eV) penetrates less than 100 nm in FeSe, while the typical thickness of a cleaved FeSe single crystal is $\sim20~\rm{\mu}$m, RIXS studies of FeSe are free from signal contamination due to {\BFA} and provide a unique opportunity for measuring high-energy magnetic excitations on detwinned FeSe with high efficiency.

\begin{figure*}[htbp!]
\includegraphics[width=14cm]{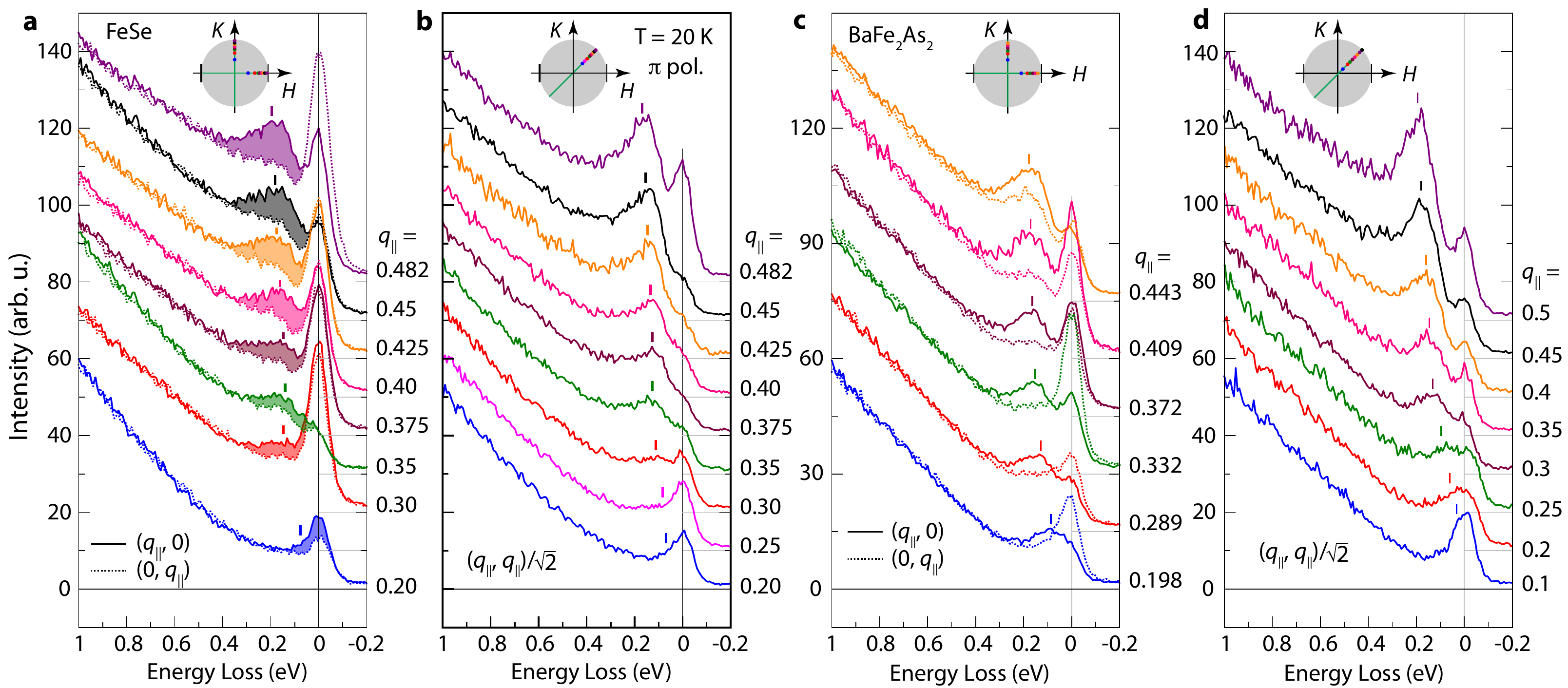}
\caption{Summary of RIXS results on detwinned FeSe and {\BFA}. {\bf a, b,} Momentum dependent RIXS spectra of FeSe along $H$ (solid lines) ({\bf a}), $K$ (dashed lines) directions ({\bf a}), and twinned FeSe along $[H, H]$ direction ({\bf b}). The colored areas mark the difference between $S_h(q_{\shortparallel})$ and $S_k(q_{\shortparallel})$. {\bf c, d,} Momentum-dependent RIXS spectra of detwinned {\BFA} along $H/K$ ({\bf c}) and  $[H, H]$ ({\bf d}) directions. The colored dots in the insets mark the momenta where spectra were collected. }
\label{fig2}
\end{figure*}

In this work, we use RIXS to measure the intrinsic spin excitations of FeSe and {\BFA} along high symmetry directions $H$, $K$ and $[H, H]$, denoted by $S_h(q_{\shortparallel})$, $S_k(q_{\shortparallel})$ and $S_{hh}(q_{\shortparallel})$, respectively (Figs. 2 and 3). To facilitate discussions,  we define the spin excitations associated with ${\bf Q}_1 = (1, 0)$ as $S_1(q, E)$ and that associated with ${\bf Q}_2=(0, 1)$ as  $S_2(q, E)$. The ratio between $S_h(q_{\shortparallel})$ and $S_k(q_{\shortparallel})$, $\psi(q_{\shortparallel})=S_h(q_{\shortparallel})/S_k(q_{\shortparallel})$, directly probes the spin-excitation unbalance between $S_1$ and $S_2$, which is commonly termed nematic spin correlations in the nematic ordering and fluctuating
region \cite{Lu14, Lu18,Liu20.1}. We denote the momentum transfer in reciprocal lattice units (r.l.u.) (See Experimental Setups in Methods).

 Our results reveal that the spin-excitation anisotropy in detwinned FeSe manifests over a large energy range up to 200 meV. It persists up to a temperature slightly above $T_{\rm s}$, before fading away at a temperature well above $T_{\rm s}$. Its comparison with the intrinsic spin-wave anisotropy of {\BFA} establishes strong nematic spin correlations in both energy scale and amplitude in FeSe. This strong spin-excitation anisotropy establishes a direct connection with the nematic phase, suggesting that the nematic order is primarily spin driven because the energy range of the nematic-phase-induced spin excitation anisotropy is much larger than that of orbital order \cite{yi2019, Baek15, anna15}.
Furthermore, our results identify dispersive high-energy spin excitations that are underdamped,
which is highly peculiar for a paramagnet, but can be understood from a local-moment based model with antiferro-quadrupolar oder \cite{YuSi15}. As such, our results provide the much-needed new insights into the mechanism for the nematicity of FeSe.

Figure 1b shows a FeSe crystal prepared for RIXS measurements, which is glued onto a square-shaped {\BFA} sample with the same orientation. Uniaxial pressure is applied on {\BFA} along the tetragonal [110] direction (orthorhombic $b$ axis). Upon cooling, {\BFA} will be detwinned below $T_{\rm s}\approx138$K and generates an orthorhombic distortion $\delta=(a-b)/(a+b)=0.36\%$ that can be transferred to and thereby detwin FeSe below its structural transition at $T_{\rm s}\approx90$K. Figure 1c, d illustrates the scattering geometry, the substantial area of the first Brillouin zone accessible with Fe-$L_3$ RIXS, and calculations of the in-plane momenta $q_{\shortparallel}$.

We first carried out incident-energy dependent RIXS (energy detuning) measurements for both {\BFA} and FeSe around their resonating energies which are determined by x-ray absorption spectra (XAS) (Fig. 1e) \cite{SI}. Figure 1f shows a RIXS map for an unstrained FeSe sample measured at $q_{\shortparallel}=(0.3, 0.3)$ with $\pi$ polarisation and $T=20$K. While fluorescence and particle-hole excitations dominate the scattering signal above $\sim0.5$ eV, a clear intrinsic elementary excitation (Raman mode) is observed at $\sim160$ meV well separated from the fluorescence peak setting on above $\sim 200$ meV. The integrated intensity of this Raman mode (green dots in Fig. 1e) follows the XAS (red curve in Fig. 1e), indicating the cross section is enhanced near the Fe-$L_3$ edge. In addition, phonon contributions to this mode can also be excluded because of their much smaller energy scale ($\sim 40$ meV) \cite{boothroyd19}.
Taking together the consistency in energy dispersion between the excitations in Figs. 2 and 3 and those reported in previous studies \cite{kejin2013, boothroyd19, johnny21},  we can safely attribute these dispersive excitations to single spin-flip magnetic excitations.

To quantitatively determine the spin-excitation anisotropy in FeSe, we have measured the spin excitations of both detwinned FeSe and {\BFA} samples (Fig. 2, 3), and take the latter as a reference, which can be compared to previous neutron scattering studies of detwinned {\BFA} \cite{Lu18}. Before discussing the results, we illustrate
in Fig. \ref{fig1}g, h the principle for resolving inherent spin-wave anisotropy and nematic spin correlations in the first Brillouin zone \cite{SI}. In a twinned sample with anisotropic excitations, RIXS measurements generate $S_h$ = $S_k$ (black curve in the inset of Fig. 1g). Both $S_h$ and $S_k$ consist of two spin-excitation branches from  $S_1$ and $S_2$ in twin domains, where we assume $S_1$ and $S_2$ have the same energy dispersion for simplicity. For a detwinned sample with local-moment AF order, the spin waves emanate only from the AF vector $\mathbf{Q_1}$=(1, 0) ($S_2$=0), and $S_h/S_k$ determines the inherent spin-wave anisotropy of $S_1$. In a detwinned system with nematic spin correlations ($S_1\neq S_2\neq0$), $S_1$ and $S_2$ will be present along both $H$ and $K$ directions (Fig. \ref{fig2}b) with {\it different} spectral weight, for which the ratio between $S_h$ and $S_k$ reflects the nematic spin correlations \cite{SI}.

Momentum-dependent RIXS spectra of FeSe and {\BFA} collected at $T=20$ K $<T_s$ are summarized in Fig. \ref{fig2}. In both samples, highly dispersive magnetic excitations along three high-symmetry directions $H$, $K$ and $[H, H]$ are resolved.
Figure 2a (2c) displays the intrinsic spin excitations of FeSe ({\BFA}) along $H$ and $K$ directions, and Fig. 2b (2d) along $[H, H]$ directions. 
While twinned FeSe and {\BFA} are expected to show four-fold symmetric magnetic excitations \cite{leland11, kejin2013} [$S_h(q_{\shortparallel}) = S_k(q_{\shortparallel})$], we find that the detwinned samples exhibit highly anisotropic excitations with $S_h(q_{\shortparallel}) > S_k(q_{\shortparallel})$.
The significant difference between $S_h(q_{\shortparallel})$ and $S_k(q_{\shortparallel})$ of FeSe persists at all $q_{\shortparallel}$ measured (colored area in Fig. 2a), and therefore demonstrates the existence of a high energy spin-excitation anisotropy (nematic spin correlations) in the nematic state of detwinned FeSe. For {\BFA}, the spin-excitation anisotropy is a manifestation of the inherent difference in the spin-wave branches along $H$ and $K$ directions \cite{Lu18}. The spin-wave anisotropy of {\BFA} increases with $q_{\shortparallel}$ up to $0.409$ but decreases at higher $q_{\shortparallel}=0.443$ (Fig. 2c), revealing a non-monotonic momentum dependence.
In comparison, the spectral weight difference in FeSe (colored area in Fig. 2a) retains a large amplitude at an even higher $q_{\shortparallel}=0.482$.

\begin{figure}[htbp!]
\includegraphics[width=8 cm]{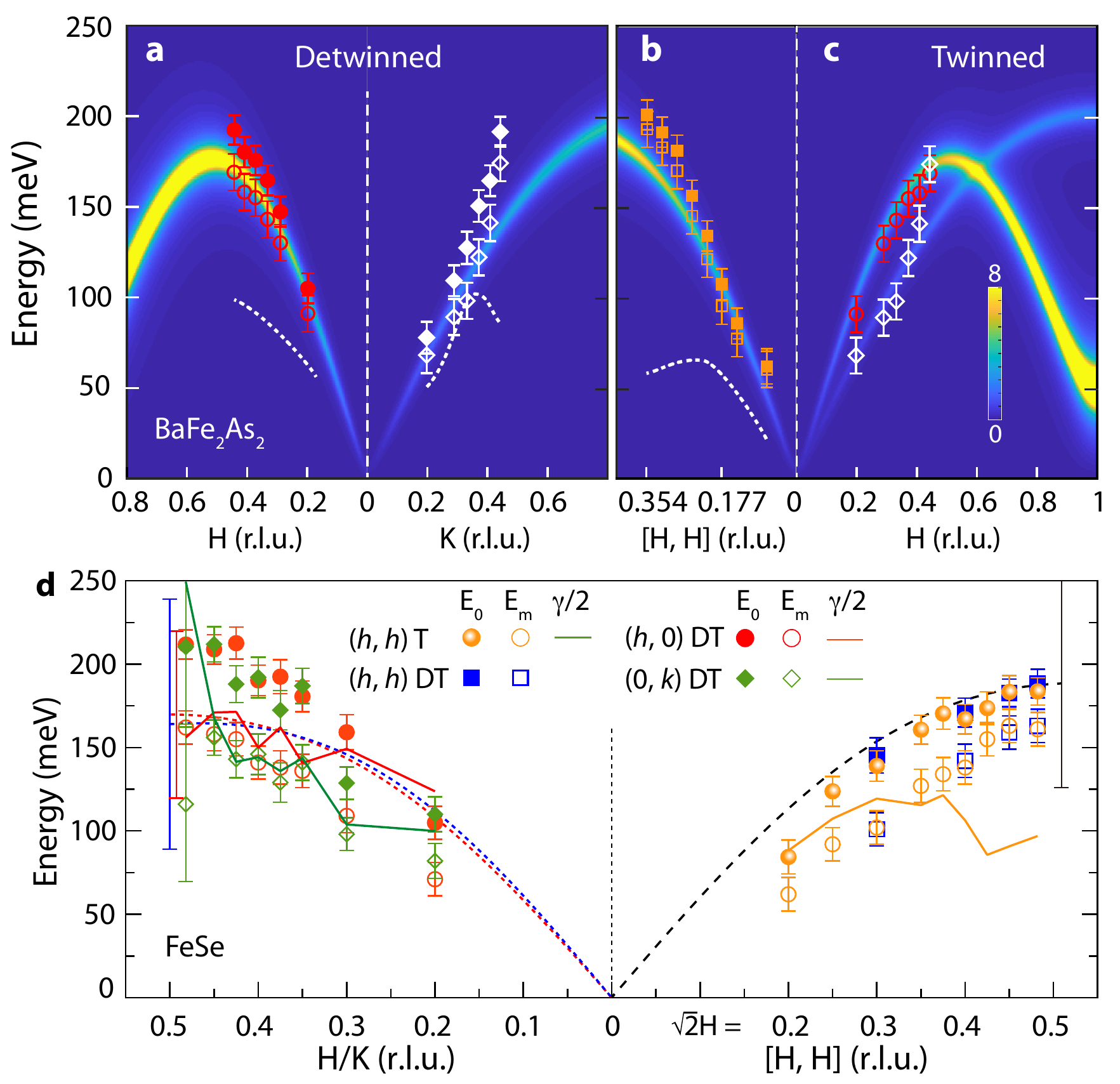}
\caption{ {\bf a-c,} Comparison between dispersions of the magnetic excitations measured with RIXS and a simulation with a Heisenberg model for {\BFA} \cite{leland11}. The solid symbols, open symbols, and white dashed curves in {\bf a} and {\bf b} are undamped energies ($E_0$), energies for the intensity maxima ($E_m$), and damping factor ($\gamma/2$) fitted from the spectra shown in Fig. 2{\bf c, d} with Eq. (1). The open symbols in {\bf c} mark the $E_m$. {\bf d,} Spin-excitation energy dispersions of FeSe obtained from the fitting of the RIXS spectra in Fig. 2{\bf a, b} with Eq. (1). The solid (open) red circles, green diamonds, orange circles, and blue squares mark the undamped energies $E_0$ (energies for intensity maxima $E_m$) of the dispersions along $H$ (detwinned), $K$ (detwinned), $[H, H]$ (twinned), and $[H, H]$ (detwinned) directions, respectively. The red, green and orange lines mark the momentum-dependent damping factor $\gamma/2$ along high symmetry directions. The black, red, and blue dashed curves are the calculated flavor-wave dispersions in the AFQ phase along $[H, H]$ and $H/K$ directions, respectively. The errorbars mark the damping factor $\gamma/2$ along the corresponding directions.}
\label{fig3}
\end{figure}

In order to achieve a quantitative characterization of the spin-excitation anisotropy in FeSe and {\BFA}, we use a general damped harmonic oscillator model \cite{monney19, boothroyd19, kejin2013, johnny17}
\begin{equation}
S(q,E)=A\,\frac{E_0}{1-{\rm e}^{-\beta E}}\frac{2\,\gamma\,E}{\left(E^2-E_0^2\right)^2+(E\,\gamma)^2},
\label{Eq1}
\end{equation}
to fit the magnetic excitations, in which $E_0$ is the undamped energy, $\gamma$ the damping factor, $\beta=\frac{1}{k_{\rm B}T}$ ($k_{\rm B}$ is Boltzmann constant) and $A$ is a fitting coefficient.   The elastic peak can be fitted with a Gaussian function, and the fluorescence contributions below $\sim1$ eV can be described with a quadratic polynomial \cite{SI}. 

The fitting results for FeSe and {\BFA}, the undamped energy dispersion ($E_0$(q)), the energies for the intensity maxima ($E_m$(q)), and the damping factor ($\gamma/2$) are summarized in Fig. 3. Figure 3a-c shows a simulation of the spin waves in {\BFA} overlaid by the energy dispersions and the damping factors. The simulation is based on an anisotropic Heisenberg $J_{1a}-J_{1b}-J_2$ model described in ref. \cite{leland11}, in which we set $L=0$ because the spin waves of {\BFA} are two dimensional, especially at the high-energies probed in our RIXS measurements. We find $E_0 > \gamma\,/2$ at all the momenta measured, which indicates that the spin waves are underdamped and far from being critically damped. 
The intrinsic spin waves of the clearly resolved two different branches $S_h(q)$ and $S_k(q)$ around $\Gamma$ is consistent with the anisotropic Heisenberg model, in which dispersive spin waves can only arise from the AF wave vector ${\bf Q_1}$ ($S_1$). The minor deviation of the branch $S_k(q)$ from the
anisotropic Heisenberg model can be attributed to the failure of the
anisotropic Heisenberg model in describing the small anisotropy of magnetic excitations at high energy,
due to the emergence of the spin excitations around ${\bf Q_2}=(0, 1)$ ($S_2$) at $E\gtrsim100$meV
observed by neutron scattering \cite{Lu18,Liu20.1}.
Figure 3d shows the energy dispersion of FeSe obtained from the fitting of the magnetic excitations shown in Fig. 2a, b. The solid symbols mark the bare energy dispersion ($E_0$) without damping effect, and the open symbols represent the energy dispersion for the intensity maxima $E_m$.

\begin{figure*}[htbp!]
\includegraphics[width=16cm]{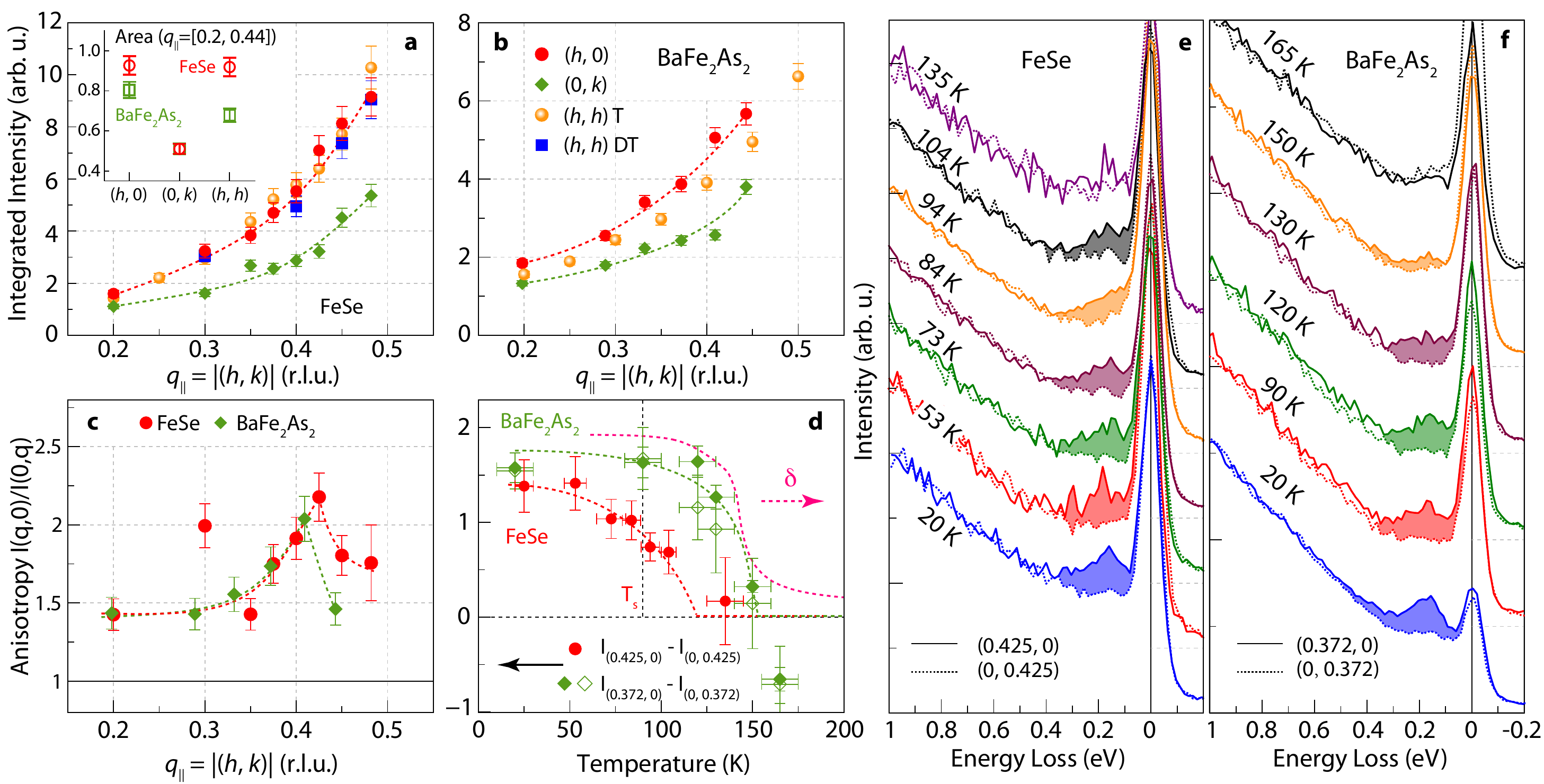}
\caption{ Anisotropic magnetic excitations in detwinned FeSe and {\BFA}. {\bf a, b,} Momentum-dependent energy-integrated intensity of magnetic excitations for FeSe and {\BFA}. The inset shows the momentum-integrated intensity of $S_{h/k/hh}$ for FeSe (red circles) and {\BFA} (green squares) in the range of $q_{\shortparallel} = [0.2, 0.44]$. {\bf c}, Spin excitation anisotropy between $S_{h}$ and $S_{k}$, defined as the ratio between the integrated intensity for $(q, 0)$ and $(0, q)$ as shown in {\bf a} and {\bf b}. {\bf d,} Temperature dependence of the spin-excitation difference between $S_h(q_{\shortparallel})$ and $S_k(q_{\shortparallel})$,  in which the data points are integrated intensity of $S_h(q_{\shortparallel})-S_k(q_{\shortparallel})$ in the energy range of [0.08, 0.4] eV as shown in {\bf e} and {\bf f}. The dashed lines in {\bf c} and {\bf d} are guides to the eye. The pink dashed curve shows the lattice distortion $\delta=(a-b)/(a+b)$ of {\BFA} under a uniaxial pressure of $\sim 20$ MPa, which reaches 0.36\% below $\sim 100$ K. The vertical black dashed line marks the $T_{\rm s}=90$ K for FeSe. {\bf e, f,} Temperature-dependent RIXS spectra for FeSe and {\BFA} measured at $q_{\shortparallel}= 0.425$ and $0.372$, respectively. The colored areas in {\bf e} and {\bf f} mark the intensity difference between $S_h(q_{\shortparallel})$ and $S_k(q_{\shortparallel})$.}
\label{fig4}
\end{figure*}

The momentum-dependent damping factors $\gamma/2$ are overall larger than that for {\BFA} but still in the underdamped regime for most of the excitations. Moreover, the damping factor for $S_{hh}$ (along $[H, H]$ direction) is smaller than for $S_{h/k}$ in both FeSe and {\BFA}, suggesting a common anisotropic damping effect in FeSC. The (anisotropic) unerdamped nature of the magnetic excitations, not inferable in
previous neutron scattering and RIXS studies on twinned samples \cite{Qwang16, boothroyd19}, which favors a local moment picture with strong electron correlation, can now be conclusively revealed and discussed.

\begin{figure*}[htbp!]
\includegraphics[width=12cm]{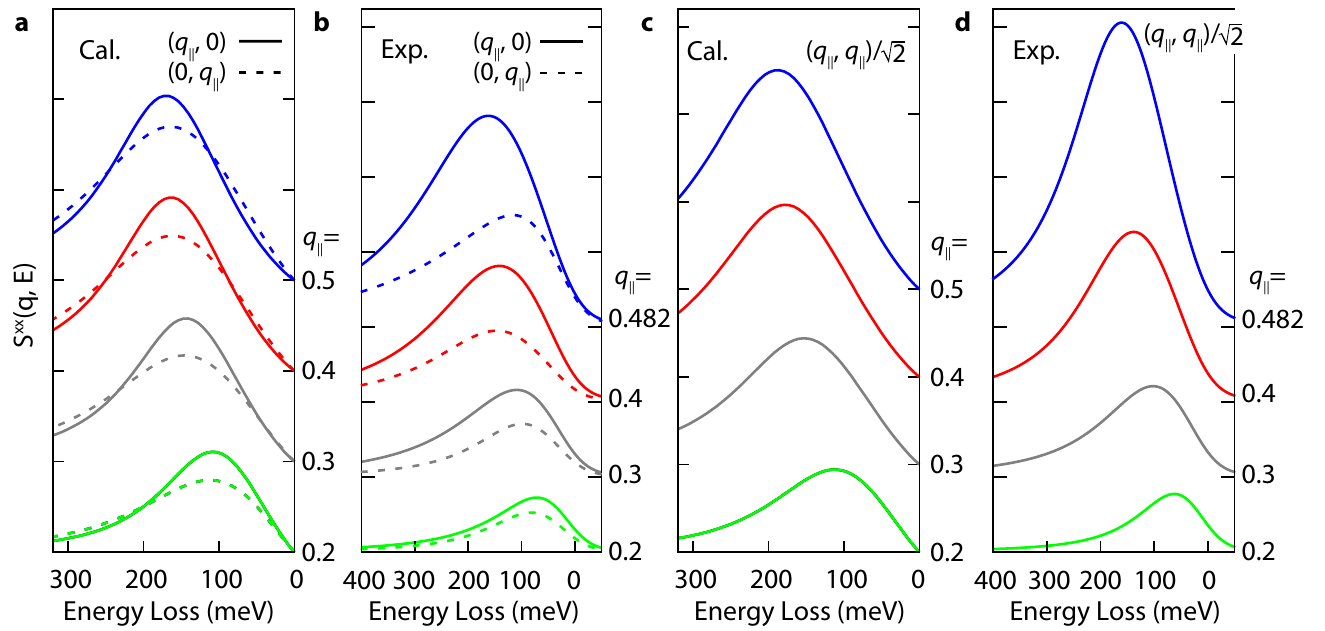}
\caption{Calculated spin excitation spectra of the AFQ phase and their comparison with the fitting curve of the experimental $S(q_{\shortparallel})$ from RIXS. {\bf a, c} Calculated spectra along the $H/K$ directions ({\bf a}), and the $[H, H]$ direction ({\bf c}). {\bf b, d} Fitting curves of the experimental $S(q_{\shortparallel})$ along the $H/K$ directions ({\bf b}) and $[H, H]$ direction ({\bf d}).
}
\label{fig5}
\end{figure*}

Since the largest momentum (0.482, 0) along $H$ is close to the zone boundary (0.5, 0), the excitation energy scale of FeSe ($E_0\sim200$ meV and $E_m\sim160$ meV) at (0.482, 0) and (0, 0.482) (consistent with that in ref. \cite{boothroyd19}) reveals a much higher band top in the first BZ than that ($\sim 120-150$ meV) observed by neutron scattering in the BZ around (1, 0) \cite{Qwang16}. This is in stark contrast to the case of {\BFA} where the dispersions measured by RIXS and neutron scattering can be consistently described with one model (Fig. 2) \cite{Lu18, kejin2013}.
We attribute this difference to the absence of stripe 
AF order in FeSe. Because of the translational symmetry of the stripe 
AF order ($\mathbf{k}=(1, 0)$) in {\BFA}, the magnetic excitations in the first BZ can be deemed as replica of the ones 
in the BZ centering at $\mathbf{Q}=(1, 0)$. However, long range 
AF order is not established in spite of the strong in-plane magnetic correlations in FeSe. 
Thus the dispersion around $\Gamma$ and (1, 0) do not have to be identical. However, how to quantitatively reconcile the neutron scattering and RIXS results is still an open question.

Figure 4a and 4b show the momentum-dependent energy-integrated intensity of the spin excitations $S_h$, $S_k$ and $S_{hh}$ for FeSe and {\BFA}, respectively. With increasing $q_{\shortparallel}$, the integrated intensities along all three high symmetry directions increase monotonically. The spin-excitation intensity for FeSe is slightly higher than {\BFA}, qualitatively consistent with previous neutron scattering results \cite{Qwang16}. Note the momentum-direction dependence of the amplitude for $S_{h/k/hh}$ is modulated by the anisotropic damping factor inherent to $S_{1,2}$, which require $S_{1,2}=0$ at $\Gamma$ \cite{leland11}.
To quantify the spin-excitation anisotropy, we plot the ratio  $I_{(q, 0)}/I_{(0, q)} = S_h/S_k$ for FeSe and {\BFA} in the same panel (Fig. 4c). It is surprising that the spin-excitation anisotropy of FeSe (reflecting the nematic spin correlation) is rather similar to the spin-wave anisotropy of {\BFA} in both amplitude and energy scale. Furthermore, at high-energy/momentum region, the anisotropy of FeSe is even larger than for {\BFA}. This prominent spin-excitation anisotropy signifies strong electronic nematicity with large energy scale ($\sim 200$ meV) and magnitude in FeSe. Moreover, the temperature dependence of the difference between $S_h$ and $S_k$ for FeSe at selected $q_{\shortparallel}=0.425$ decreases with increasing temperature, persists to a temperature ($104$ K) 20\% higher than $T_{\rm s}$ and finally drastically reduces at a temperature ($135$ K) well above $T_{\rm s}$, indicating a gradual suppression of nematic spin correlations above $T_{\rm s}$ (Figs. 4d and 4e). Since FeSe is under uniaxial strain applied from the {\BFA} substrate (pink dashed curve in Fig. 4d), $T_{\rm s}$ is no longer well defined.  For this reason, it is not surprising that the spin-excitation anisotropy disappears at a temperature above the zero pressure $T_{\rm s}$. This temperature dependence is similar to that for {\BFA} as shown in Figs. 4d and 4f, in which the anisotropy at $q_{\shortparallel}=0.372$ persists at a temperature slightly higher than $T_{\rm N}$ but vanishes at $T = 165$ K.

Previous measurements of the spin dynamics 
in the detwinned FeSe were constrained to low energies ($\lessapprox$ 10 meV) \cite{Chen19},
and were unable to discriminate between the different scenarios for the nematicity in FeSe. 
Our present measurements over a large energy window enables the discovery that the detwinned FeSe 
harbors high-energy (up to $200$ meV) anisotropic spin excitations that are dispersive and underdamped. 
This surprising finding is at variance with the itinerant mechanism for the nematicity in FeSe. In the itinerant picture, the leading contribution to the spin excitation spectrum comes from a two-particle process, {\it i.e.} a convolution of single electron and single hole excitations. Because of this particle-hole nature of the spin excitations, damping already appears in the leading-order contribution to the spectrum. This Landau damping means that the leading-order contribution to the spin excitations is, generically, already overdamped; higher-order processes further enhance the damping. An itinerant model calculation indeed shows that the spin excitations at such the pertinent (high) energies are highly overdamped \cite{Kreisel18}. Our work is expected to motivate further studies on this issue within the itinerant description.

Instead, the dispersive and underdamped nature of the spin excitations point to a local-moment starting point 
to describe the nematicity. The notion that had emerged from the iron pnictides \cite{dai2012} is that the majority of the spin degree of freedom resides in quasi-localized moments, especially at high energies relevant to the RIXS measurements.
This has led us to consider a generalized bilinear-biquadratic model on a square lattice with local moments \cite{YuSi15}. In such models, the leading-order contributions to the spin excitation spectrum is a one-particle process. Because only sub-leading contributions that involve more-than-one-particle processes contribute to damping, the high-energy spin excitations are underdamped. More specifically, we have calculated the spin dynamics 
in the proposed picture for the nematicity based on a $(1, 0)$ antiferroquadrupolar state using the flavor wave method \cite{YuSi15}. This approach treats the spin dipolar and quadrupolar excitations on equal footing, which are the bosonic flavor waves. Consistent with the physical picture, the spin excitation spectrum is dispersive and underdamped;
the details are 
given in the Supplementary Information.  In Fig. 5, we show that 
the calculated spin dynamics provide a good understanding of the experimental data.
Thus, our results provide evidence for a local-moment-based picture for the nematicity of FeSe.

The rich properties of the Fe-based superconductors in general and nematic FeSe in particular have been addressed from both weak \cite{Fernandes14, Brian15, Fanfarillo2016} and strong \cite{NatRevMat16, FWang15, YuSi15} coupling perspectives, with each having had successes.
Our work brings about a hitherto-unknown and remarkable feature of universality across the Fe-based superconductors.
For {\BFA} with collinear AF order, the anisotropy can be readily interpreted 
in the picture of intrinsic spin waves \cite{Lu18, Liu20.1}.
FeSe, by contrast, lacks long-range magnetic order even though it is nematic. 
The local-moment-based understanding, as dictated by the qualitative similarities 
in the high energy spin excitations between FeSe and {\BFA}, suggest that the largest spin spectral weight in both the iron chalcogenides and pnictides
is associated with the incoherent electronic excitations induced by the underlying electron correlations.
As such, our work not only allows for discriminating the proposed mechanisms for 
 the nematicity of FeSe but also points to a unified understanding of the correlation physics \cite{johnny19} across the seemingly
distinct classes of Fe-based superconductors.



{\bf Acknowledgement}

The work at Beijing Normal University is supported by the National Natural Science Foundation of China (Grant No. 11922402 and 11734002). The RIXS experiments were carried out at the ADRESS beamline of the Swiss Light Source at the Paul Scherrer Institut. The work at PSI is supported by the Swiss National Science Foundation through project no. 200021\_178867, and the Sinergia network Mott Physics Beyond the Heisenberg Model (MPBH)  (project numbers CRSII2 160765/1 and CRSII2 141962).
The work at Renmin University was supported by the Ministry of Science and Technology of China, National Program on Key Research Project Grant No.2016YFA0300504 and Research Funds of Remnin University of China Grant No. 18XNLG24 (R.Y.)
The experimental work at Rice university is supported by the U.S. Department of Energy, BES under Grant No. DE-SC0012311 (P.D.). The single-crystal synthesis work at Rice is supported by Robert A. Welch Foundation Grant No. C-1839 (P.D.). The theoretical work at Rice was supported by
the U.S. Department of Energy, Office of Science, Basic Energy Sciences, under Award No. DE-SC0018197,
and the computational part by
the Robert A.\ Welch Foundation Grant No.\ C-1411 (Q.S.).
Work at Los Alamos was carried out under the auspices of the U.S. DOE NNSA under Contract No. 89233218CNA000001.
It was supported by LANL LDRD Program and in part by the Center for Integrated Nanotechnologies, a U.S. DOE BES user facility.
Q.S. acknowledges the hospitality of the Aspen Center for Physics,
which is supported by NSF grant No. PHY-1607611.

{\bf Methods}

{\bf Sample preparation} The high-quality {\BFA} and FeSe single crystals used in the present study were grown using self-flux and chemical vapor transport method, respectively. The {\BFA} single crystals were oriented using a Laue camera and cut along tetragonal [110] and [1-10] directions using a high-precision wire saw (WS-25). The direction of the self-cleaved edges of FeSe single crystals were also determined using a Laue camera. The well-cut {\BFA} crystals, with typical size $5$mm$\times4.3$mm$\times0.5$mm, were pre-cleaved before the final preparation. For RIXS measurements of {\BFA}, we put a ceramic top-post onto the upper surface of {\BFA} for in-situ cleaving. For RIXS measurements of FeSe, we glue thin FeSe crystals onto the upper surface of {\BFA} along the same direction using epoxy Stycast 1266 and a small ceramic top-post onto the surface of FeSe. The prepared crystals with the posts were inserted into the slot of the uniaxial-pressure devices, which were mounted on a modified copper sample holder of the RIXS spectrometer (See Fig. \ref{fig1}(b)).

{\bf Experimental setups} The RIXS and XAS measurements were performed with the RIXS spectrometer at ADRESS beamline of Swiss Light Source at the Paul Scherer Institut \cite{ADRESS, SAXES}. The beam size at the sample position is $\sim10\times50\rm{\mu}$m$^2$. All the measurements shown in the main text were collected using linear horizontal (LH) polarization (electric field vector of the incident photons lying within the horizontal scattering plane), denoted as $\pi$ polarization. The RIXS spectra were collected with a grazing-incidence configuration, as shown in Fig. \ref{fig1}(d). The scattering angle was set to $2\theta_s=130^{\circ}$, by which a substantial area of the first Brillouin zone is accessible [gray circle in Fig. 1(d)]. The measurements were performed along high-symmetry directions $H/K$ and $[H, H]$ in orthorhombic notation. The total energy resolution for the RIXS measurements was set to 80 meV. The in-plane momentum $q_{\shortparallel}$ can be tuned continuously by rotating the sample and thereby changing the angle ($\delta$). Before measurements, the sample holder was inserted into the manipulator head and the top-post was removed by cleaving at low temperature ($\sim20$K) and ultra-high vacuum ($<10^{-10}$ mbar). Through the experiments, we define reciprocal lattice unit as $(q_xa_\mathrm{o}/2\pi, q_yb_\mathrm{o}/2\pi, q_zc/2\pi)$ with $a_\mathrm{o}\approx 5.334${\AA}, $b_\mathrm{o}\approx 5.308${\AA} and $c\approx 5.486${\AA}. The FeSe tri-layer height is $d\approx 5.5${\AA}. The orthorhombic lattice distortion of FeSe is $\delta=(a_\mathrm{o}-b_\mathrm{o})/(a_\mathrm{o}+b_\mathrm{o})\approx 0.27\%$ at a temperature well below $T_{\rm s}$.

{\bf Author contributions}

X.L. conceived this project and developed the detwinning strategy. X.L. and T.S. wrote the beamtime proposals and coordinated together the experiments as well as all other project phases. X.L., W.Z., Y.T., E.P., R.L., Z.T., and T.S. carried out the RIXS experiments with the support of V.S.. P.L., R.L., and Z.T. prepared the {\BFA} single crystals. T.C. and P.D. provided the FeSe single crystals. R.Y. and Q.S. carried out theoretical and computational analyses.
X.L., P.D., and T.S. wrote the manuscript with inputs from R.Y. and Q.S.. All authors made comments.


\begin{thebibliography}{}

\bibitem{fradkin15} Fradkin, E., Kivelson, S. A., \& Tranquada, J. M., Colloquium: Theory of intertwined orders in high temperature superconductors. Rev. Mod. Phys. {\bf 87}, 457 (2015).

\bibitem{JHChu2010} Chu, J.-H. {\it et al.} In-plane resistivity anisotropy in an underdoped iron arsenide superconductor. Science {\bf 329}, 824-826 (2010).

\bibitem{yi2011} Yi, M. {\it et al.} Symmetry-breaking orbital anisotropy observed for detwinned {\BFCA} above the spin density wave transition. Proc. Natl Acad. Sci. USA 108, 6878-6883 (2011).

\bibitem{JHChu2012} Chu, J.-H., Kuo, H.-H., Analytis, J. G., \& Fisher, I. R.
Divergent Nematic Susceptibility in an Iron Arsenide
Superconductor, Science {\bf 337}, 710 (2012).

\bibitem{Fernandes14} Fernandes, R. M., Chubukov, A. V. \& Schmalian, J. What drives nematic order in iron-based superconductors? Nat. Phys {\bf 10}, 97-104 (2014).

\bibitem{NatRevMat16} Si, Q., Yu, R. \& Abrahams, E, Hightemperature superconductivity in iron pnictides and chalcogenides,
Nat. Rev. Mater. {\bf 1}, 16017 (2016).

\bibitem{Kuo2016} Kuo, H.-H., Chu, J.-H., Palmstrom, J. C., Kivelson, S. A., \& Fisher, I. R. Ubiquitous signatures of nematic quantum criticality in optimally doped Fe-based superconductors. Science {\bf 352}, 958-962 (2016).

\bibitem{bohmer16} B\"{o}hmer, A. E. \& Meingast, C. Electronic nematic susceptibility of iron-based superconductors. C. R. Physique {\bf 17} 90-112 (2016).

\bibitem{bohmer18} B\"{o}hmer, A. E. \& Kreisel, A., Nematicity, magnetism and superconductivity
in FeSe. J. Phys.: Condens. Matter {\bf 30}, 023001 (2018).

\bibitem{senthil} Metlitski, M. A. {\it et al.} Cooper pairing in non-Fermi liquids. Phys. Rev. B {\bf 91}, 115111(2015).

\bibitem{kivelson} Lederer, S. {\it et al.} Enhancement of Superconductivity near a Nematic Quantum Critical Point. Phys. Rev. Lett. {\bf 114}, 097001 (2015).

\bibitem{paglione20} Eckberg, C. {it et al.} Sixfold enhancement of superconductivity in a tunable electronic nematic system. Nat. Phys. {\bf 16}, 346 (2020).

\bibitem{Hsu} Hsu, F. C. {\it et al.} Superconductivity in the PbO-type structure $\alpha$-FeSe.
Proc. Natl Acad. Sci. USA {\bf 105}, 14262 (2008).

\bibitem{McQueen09} McQueen, T. M. {\it et al}. Tetragonal-to-orthorhombic structural phase transition at 90K in the superconductor Fe$_{1.01}$Se. Phys. Rev. Lett. {\bf 103}, 057002 (2009).

\bibitem{tanatar16} Tanatar, M. A. {et al.}, Origin of resistivity anisotropy in the nematic phase of FeSe, Phys. Rev. Lett. {\bf 117}, 127001 (2016).

\bibitem{Coldea18} Coldea, A. \& Watson M. D., The key ingredients of the electronic
structure of FeSe. Annu. Rev. Condens. Matter Phys. {\bf 9}, 125-146 (2018).

\bibitem{Liu2018} Liu, D. F. {\it et al.} Orbital origin of extremely anisotropic superconducting gap in nematic phase of FeSe superconductor. Phys. Rev. X {\bf 8}, 031033 (2018).

\bibitem{Hashimoto2018} Hashimoto, T. {\it et al.} Superconducting gap anisotropy sensitive to nematic domains in FeSe. Nat. Commun. {\bf 9}, 282 (2018).

\bibitem{Rhodes2018} Rhodes, L. C. {\it et al.} Scaling of the superconducting gap with orbital character in FeSe. Phys. Rev. B {\bf 98}, 180503(R) (2018).

\bibitem{sprau17} Sprau, P. O. {\it et al.} Discovery of orbital-selective Cooper pairing in FeSe, Science {\bf 357}, 75-80 (2017).

\bibitem{CCLee2009} Lee, C. C., Yin, W.-G., \& Ku, W., Ferro-orbital order and strong magnetic anisotropy in the parent compounds of iron-pnictide superconductors, Phys. Rev. Lett. {\bf 103},
267001 (2009).

\bibitem{yi2019} Yi, M. {\it et al.} The nematic energy scale and the missing electron pocket in FeSe, Phys. Rev. X {\bf 9}, 041049 (2019).

\bibitem{Baek15} Baek, S.-H. {\it et al.}
Orbital-driven nematicity in FeSe. Nat. Mater. {\bf 14}, 210 (2015).

\bibitem{anna15} B\"{o}hmer, A. E. {\it et al.} Origin of the tetragonal-to-orthorhombic phase transition in FeSe: A combined thermodynamic and NMR study of nematicity. Phys. Rev. Lett.  {\bf 114}, 027001 (2015).

\bibitem{Yamakawa16} Yamakawa, Y., Onari, S., \& Kontani, H. Nematicity and magnetism in FeSe
and other families of Fe-based superconductors, Phys. Rev. X {\bf 6}, 021032 (2016).

\bibitem{Onari16} Onari, S., Yamakawa, Y., \& Kontani, H.
Sign-reversing orbital polarization in the nematic phase of
FeSe due to the $C_2$ symmetry breaking in the self-energy,
Phys. Rev. Lett. {\bf 116}, 227001 (2016).

\bibitem{wang16} Wang, Q. {\it et al.} Strong interplay between stripe spin fluctuations, nematicity and superconductivity in FeSe. Nat. Mater. {\bf 15}, 159 (2016).

\bibitem{Qwang16} Wang, Q. {\it et al.} Magnetic ground state of FeSe. Nat. Commun. {\bf 7}, 12182 (2016).

\bibitem{MWMa17} Ma, M. W. {\it et al.} Prominent role of spin-orbit coupling in FeSe
revealed by inelastic neutron scattering, Phys. Rev. X {\bf 7}, 021025 (2017).

\bibitem{FWang15} Wang, F., Kivelson, S. \& Lee, D.-H. Nematicity and quantum paramagnetism in FeSe.
Nat. Phys. {\bf 11}, 959-963 (2015).

\bibitem{YuSi15} Yu, R. \& Si, Q. Antiferroquadrupolar and Ising-nematic orders of a frustrated
bilinear-biquadratic Heisenberg model and implications for the magnetism of FeSe.
 Phys. Rev. Lett. {\bf 115}, 116401 (2015).

\bibitem{Glasbrenner15} Glasbrenner, J. K. {\it et al.} Effect of magnetic frustration on nematicity and superconductivity in iron chalcogenides. Nat. Phys. {\bf 11}, 953 (2015).

\bibitem{She17} She, J.-H., Lawler, M. J., \& Kim, E.-A., Quantum spin liquid intertwining nematic and superconducting order in FeSe, Phys. Rev. Lett. {\bf 121}, 237002 (2018).

\bibitem{Lu14} Lu, X. {\it et al.} Nematic spin correlations in the tetragonal state of uniaxial-strained BaFe$_{2-x}$Ni$_{x}$As$_2$, Science {\bf 345}, 657 (2014).

\bibitem{Chen19} Chen, T. {\it et al.}, Anisotropic spin fluctuations in detwinned FeSe. Nat. Mater. {\bf 18}, 709-716 (2019).

\bibitem{Yu14} Yu, R. Zhu, J.-X. \& Si, Q., Orbital-selective superconductivity, gap anisotropy and spin resonance excitations in a multiorbital $t$-$J_1$-$J_2$ model for iron pnictides. Phys. Rev. B {\bf 89}, 024509 (2014).

\bibitem{Fanfarillo2016} Fanfarillo, L. {\it et al.} Orbital-dependent Fermi surface shrinking as a fingerprint of nematicity in FeSe. Phys. Rev. B {\bf 94}, 155138 (2016).

\bibitem{kejin2013} Zhou, K. {\it et al.}, Persistent high-energy spin excitations in iron-pnictide superconductors. Nat. Commun. {\bf 4}, 1470 (2013).

\bibitem{johnny17} Pelliciari {\it et al.}, Local and collective magnetism of EuFe$_2$As$_2$. Phys.
Rev. B {\bf 95}, 115152 (2017).

\bibitem{monney19} Garcia, F. A. {\it et al.}, Anisotropic magnetic excitations and incipient N{\'e}el order in Ba(Fe$_{1-x}$Mn$_x)_2$As$_2$. Phys. Rev. B {\bf 99}, 115118 (2019).

\bibitem{johnny19} Pelliciari, J. {\it et al.}, Reciprocity between local moments and collective magnetic excitations in the phase diagram of BaFe$_2$(As$_{1-x}$P$_x$)$_2$. Communications Physics {\bf 2}, 139 (2019).

\bibitem{boothroyd19} Rahn, M. C. {\it et al.}, Paramagnon dispersion in $\beta$-FeSe observed by Fe $L$-edge resonant inelastic x-ray scattering. Phys. Rev. B {\bf 99}, 014505 (2019).

\bibitem{johnny21} Pelliciari, J. {\it et al.}, Evolution of spin excitations from bulk to monolayer FeSe, Nature Communications {\bf 12}, 3122 (2021).


\bibitem{rixs_rmp} Ament, L. J. P. {\it et al.}, Resonant inelastic x-ray scattering studies of elementary excitations. Rev. Mod. Phys. {\bf 83}, 705 (2011).

\bibitem{thorsten12} Schlappa, J. {\it et al.}, Spin-orbital separation in the quasi-one-dimensional Mott insulator Sr$_2$CuO$_3$. Nature {\bf 485}, 82-85 (2012).

\bibitem{peng17} Peng, Y. Y. {\it et al.}, Influence of apical oxygen on the extent of in-plane exchange interaction in cuprate superconductors. Nat. Phys. {\bf 13}, 1201 (2017).

\bibitem{chunjing_prx} Jia, C. {\it et al.}, Using RIXS to uncover elementary charge and spin excitations. Phys. Rev. X {\bf 6}, 021020 (2016).

\bibitem{weisheng18} Hepting, M. {\it et al.}, Three-dimensional collective charge excitations in electron-doped copper oxide superconductors. Nature {\bf 563}, 374-378 (2018).


\bibitem{Lu18} Lu, X. {\it et al.}, Spin waves in detwinned BaFe$_{2}$As$_{2}$, Phys. Rev. Lett. {\bf 121}, 067002 (2018).

\bibitem{Liu20.1} Liu, C. {\it et al.}, Anisotropic magnetic excitations of a frustrated bilinear-biquadratic spin
model -- Implications for spin waves of detwinned iron pnictides. Phys. Rev. B {\bf 101}, 024510 (2020).

\bibitem{SI} See the supplemental materials for details.

\bibitem{leland11} Harriger, L. W. {\it et al.}, Nematic spin fluid in the tetragonal phase of {\BFA}, Phys. Rev. B {\bf 84}, 054544 (2011).

\bibitem{Tam20} Tam, D. W. {\it et al.} Orbital selective spin waves in detwinned NaFeAs, Phys. Rev. B {\bf 102}, 054430 (2020).

\bibitem{ZP14} Yin, Z. P. Haule, K. and Kotliar, G. Spin dynamics and orbital-antiphase pairing symmetry in iron-based superconductors. Nat. Phys. {\bf 10}, 845-850 (2014).

\bibitem{Kreisel18} Kreisel, A. Andersen, B. M. \& Hirschfeld, P. J. Itinerant approach to magnetic neutron scattering of FeSe: effect of orbital selectivity. Phys. Rev. B {\bf 98}, 214518 (2019).

\bibitem{dai2012} Dai, P. Hu, J. \& Dagotto, E. Magnetism and its microscopic origin in iron-based high-temperature superconductors. Nat. Phys. {\bf 8}, 709 (2012).

\bibitem{Brian15} Mukherjee, S. Kreisel, A. Hirschfeld, P. J. \& Andersen, B. M. Model of Electronic Structure and Superconductivity in Orbitally Ordered FeSe. Phys. Rev. Lett. {\bf 115}, 026402 (2015).

\bibitem{ADRESS} Strocov, V. N. {\it et al.}, High-resolution soft X-ray beamline ADRESS at the Swiss Light Source for resonant inelastic X-ray scattering and angle-resolved photoelectron spectroscopies. Journal of synchrotron radiation 17, 631-643. (2010).

\bibitem{SAXES} Ghiringhelli G. {\it et al.}, A high resolution spectrometer for resonant x-ray emission in the 400-1600eV energy range. Review of Scientific Instruments 77, 113108 (2006).

\end{thebibliography}
\end{document}